\documentclass[english,aps,amssymb,prl,twocolumn, showpacs]{revtex4-1}
\pdfoutput=1
\usepackage[T1]{fontenc}
\usepackage[latin9]{inputenc}
\usepackage{amsmath}
\usepackage{graphicx}
\usepackage{amssymb}
\usepackage{esint}

\makeatletter
\@ifundefined{textcolor}{}
{% \definecolor{BLACK}{gray}{0}
 \definecolor{WHITE}{gray}{1}
 \definecolor{RED}{rgb}{1,0,0}
 \definecolor{GREEN}{rgb}{0,1,0}
 \definecolor{BLUE}{rgb}{0,0,1}
 \definecolor{CYAN}{cmyk}{1,0,0,0}
 \definecolor{MAGENTA}{cmyk}{0,1,0,0}
 \definecolor{YELLOW}{cmyk}{0,0,1,0}
 }

\renewcommand{\phi}{\varphi}
\renewcommand{\epsilon}{\varepsilon}

\renewcommand{\vec}[1]{{\bf #1}}
\newcommand{Û}{$}
\usepackage{babel}

\begin{document}

\title {Majorana states and devices in magnetic structures }

\author{Teemu Ojanen}
\email[Correspondence to ]{teemuo@boojum.hut.fi}
%\author{Takuya Kitagawa$^2$}
\affiliation{Low Temperature Laboratory (OVLL), Aalto University, P.~O.~Box 15100,
FI-00076 AALTO, Finland }
%\affiliation{$^2$Physics Department, Harvard University, Cambridge, Massachusetts 02138, USA}
\date{\today}
\begin{abstract}
The pursuit for Majorana fermions is one of the top priorities in condensed matter physics at the moment. In this work we propose a new method of fabricating  Majorana Josephson devices in systems with a weak or no spin-orbit coupling and without external magnetic fields. Our proposal is based on curved semiconductor wires in the proximity of superconducting elements and a small number of nanomagnets. With this method it is possible to fabricate devices that are not feasible by employing straight topological wire segments. The proposed method is naturally scalable and opens up a possibility for a systematic fabrication of arrays of Majorana states where a pair of Majorana states is obtained from a single magnet.

\end{abstract}
\maketitle
\bigskip{}

\emph{Introduction}-- The search for Majorana fermions in condensed-matter systems is advancing on many fronts.  Recent efforts have increasingly focused on topological insulator and superconductor systems \cite{kane2, qi}. First step to this direction was taken by Kitaev by illustrating the possibility of Majorana states in a conceptually simple $p$-wave chain of spinless fermions \cite{kitaev1}.  Fu and Kane put forward a concrete proposal to realize Majorana states in topological insulator structures proximity coupled to superconductors \cite{fu}.  That proposal was followed by the work of Lutchyn \emph{et al.} \cite{lutchyn} and Oreg \emph{et al.} \cite{oreg} who discovered that nanowires with a strong spin-orbit coupling in the proximity of a superconductor could support Majorana states above critical magnetic fields. These works provided a major boost to experimental investigations of Majorana physics in nanowires \cite{mourik,das,rokhinson, deng}. Experimental signatures of Majorana states in these systems are broadly consistent with current theoretical understanding but insufficient in settling the existence of Majorana states conclusively. Majorana bound states (MBS) obey non-Abelian braiding statistics \cite{moore, volovik,read, ivanov} which is a prerequisite for topological quantum computation \cite{nayak,kitaev2,kitaev3, alicea2}. Recent theoretical and experimental breakthroughs have brought these considerations closer to realization. 

Systems combining magnetism and superconductivity have been considered as a promising route to topological superconductivity and Majorana states. A lattice of magnetic impurity sites on top of $s$-wave superconductors could provide very clean and controlled system to study MBS \cite{np,choy}. Semiconductor nanowires in the proximity of arrays of nanomagnets also give rise to topological properties similar to those in spin-orbit coupled wires \cite{klinovaja2, klinovaja,kjaergaard}. The physical mechanism of topological superconductivity in magnetic systems and spin-orbit coupled systems rely on lifting the spin degeneracy of charge carriers leading to an effective $p$-wave pairing in the low energy corner.     
\begin{figure}
\centering
\includegraphics[width=0.9\columnwidth, clip=true]{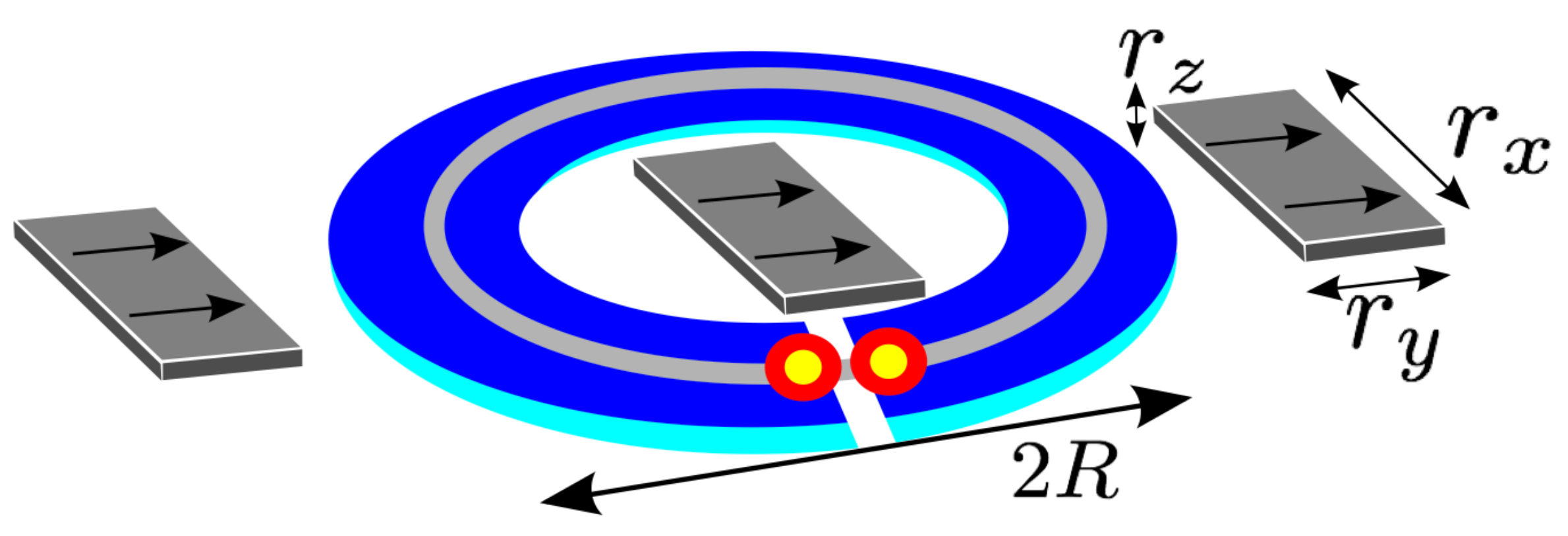}
\caption{Architecture for topological superconducting wires supporting Majorana bound states consisting of a semiconductor loop in contact with a superconductor with a weak link. The loop is located in the vicinity of three permanent magnets (arrows indicate the direction of magnetization) driving the system to the topological phase. }
\label{sche}
\end{figure}

In this work we put forward a magnetic realization of Majorana devices based on curved semiconductor wires and loops. Realizing these systems requires fabricating semiconductor loops that are made superconducting through a proximity effect and placing permanent magnets in the close vicinity, as shown in Fig.~\ref{sche}.  The loop, having a diameter of the order of one micron, and the magnets made of Fe or Co based materials can be fabricated by current technology. Considered systems have important properties compared to the systems based on straight segments of Rashba wires. The topological phase and MBS can be achieved in setups with no or weak spin-orbit coupling and it is possible to maintain the nontrivial phase without applying external magnetic field. In addition, It is also possible to fabricate devices that are not feasible by employing straight Rashba wires such as SQUID loops with only two MBS or recently introduced topological $\pi$-junctions \cite{ojanen1}.  Compared to previous studies of magnetic Majorana systems, our work takes the first concrete step towards feasible devices. Importantly, a single pair of MBS can be achieved with only a small number of magnets so large arrays are not required. However, with an array of magnets it is possible to realize a pair of MBS per magnet which enables a realization of Majorana arrays.

\emph{Studied model}-- The Hamiltonian of a one-dimensional wire in the presence of superconductivity and magnetic fields is modelled by $\mathcal{H}=\frac{1}{2}\int dx \Psi^\dagger H\Psi$, where
\begin{align}\label{h3}
H=&\left( - \frac{\hbar^2\partial_s^2}{2m}-\mu \right)\tau_z+\vec{B}\cdot\vec{\sigma}+\vec{\Delta}\cdot\vec{\tau},
\end{align}
where $\sigma_i$ and $\tau_i$ are Pauli matrices in the spin and the Nambu space, $\vec{B}=(B_x, B_y, B_z)$ represents the Zeeman splitting due to the magnetic field and the last term arises from the proximity-induced superconducting pairing with $\vec{\Delta}=\Delta\left(\mathrm{cos}\, \phi(s),\mathrm{sin}\,\phi(s),0\right)$.  The basis spinors are given by  $\Psi=(\psi_{\uparrow}, \psi_{\downarrow},\psi^{\dagger}_{\downarrow},-\psi^{\dagger}_{\uparrow} )^T$, the pairing amplitude and the superconducting phase are denoted by $\Delta$ and $\phi$ and  $\mu$ represents chemical potential of electrons. The kinetic energy of electrons is given by the usual expression $ - \frac{\hbar^2\partial_s^2}{2m}$, where $m$ is the effective mass and $\partial_s$ denotes the derivative of the wire coordinate. 

\begin{figure}
%\centering
%\includegraphics[width=0.35\columnwidth, clip=true]{maj1}
\includegraphics[width=0.4\columnwidth, clip=true]{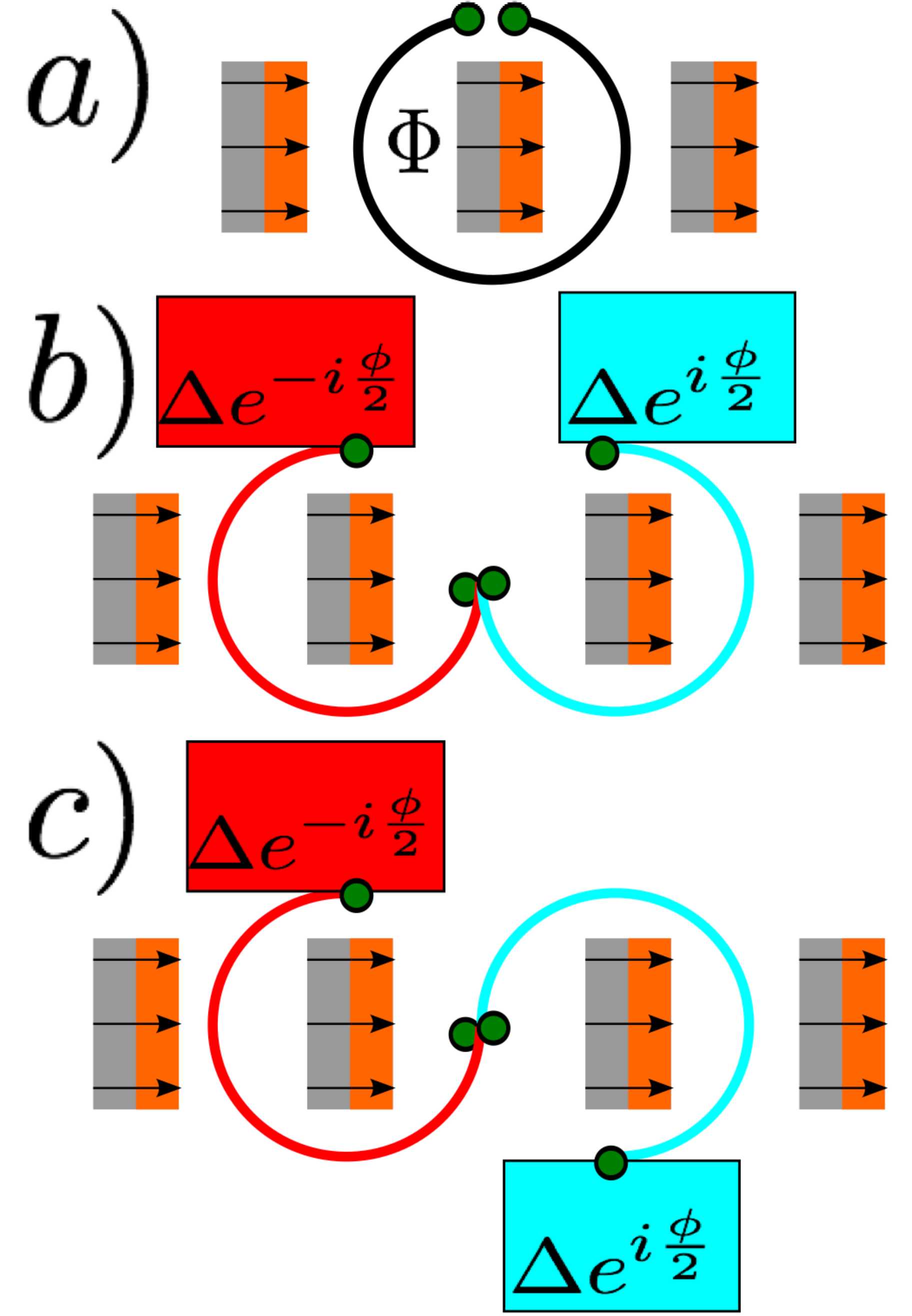}
\includegraphics[width=0.18\columnwidth, clip=true]{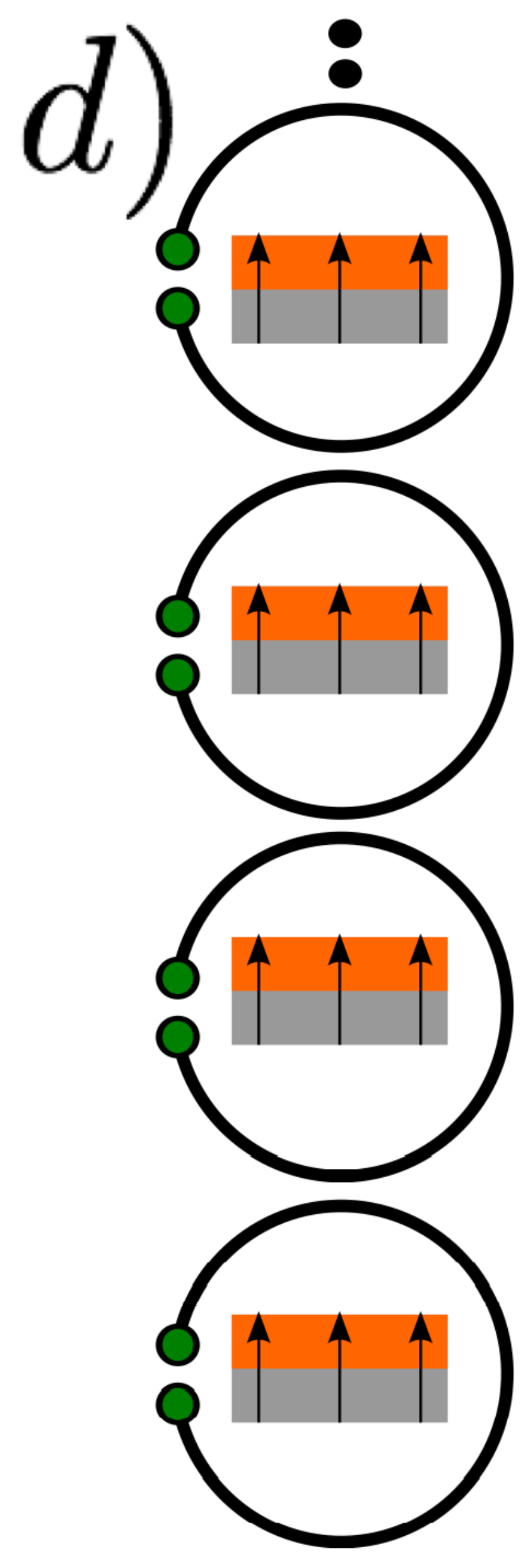}
\caption{Possible magnetic Majorana setups, MBS are schematically shown as green dots.  a): RF SQUID geometry. b): Josephson junction connecting two reservoirs.  c): Topological $\pi$-Josephson junction. d): Possible geometry for a Majorana array with an efficient MBS/magnet fraction.  }
\label{dev}
\end{figure}

In the following we are interested in geometries depicted in Figs.~\ref{sche} and \ref{dev}, where the wire consists of circular arcs in the presence of a field of bar magnets.  Even under the assumption of a homogeneous magnetization of magnets the field experienced by  electrons in the wire is strongly inhomogeneous, rendering an analytic calculation of the spectrum intractable. However, a qualitative understanding of the topological properties can be achieved along the lines of Ref.~\cite{kjaergaard}. When a particle is moving in a uniformly rotating planar magnetic field of a constant magnitude, the Hamiltonian is unitary equivalent with a one describing particle experiencing a Rashba spin-orbit coupling and a constant field. The effective Rashba constant is given by $\alpha=\frac{\hbar^2\theta'}{2m}$, where $\theta'$ is the rate of change of the field angle as a function of the wire coordinate \cite{kjaergaard}. In the geometry of Fig.~\ref{sche} the field makes two rotations along the loop, so an average effective Rashba coupling can be estimated by $\alpha=\frac{\hbar^2}{mR}$, where $R$ is the radius of the loop. When $R$ is  a few hundred nanometers, $\alpha$ could become comparable to the values of the intrinsic Rashba coupling in strongly coupled wires.  As  established in Refs.~\cite{lutchyn, oreg},  Rashba wires enter to the topologically nontrivial phase when the Zeeman field exceeds a critical value  $B_c=\sqrt{\Delta^2+ \mu^2}$ \cite{lutchyn, oreg}. These considerations, while not directly applicable to the case of inhomogeneous magnetic fields, suggest that the setup in Fig.~\ref{sche} enter to the topological phase at sufficiently strong Zeeman fields. 

 \begin{figure}
%\centering
\includegraphics[width=0.45\columnwidth, clip=true]{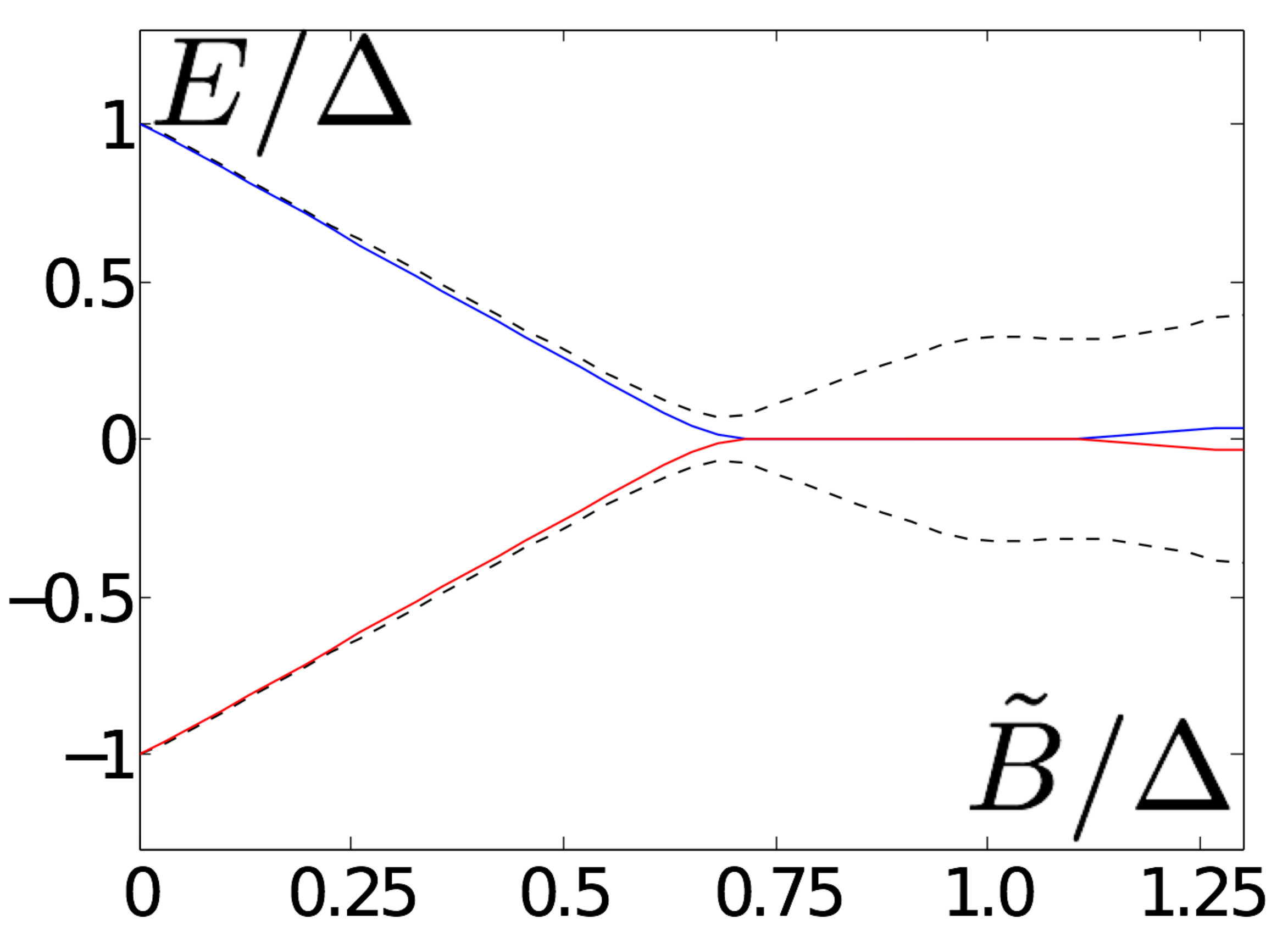}
\includegraphics[width=0.45\columnwidth, clip=true]{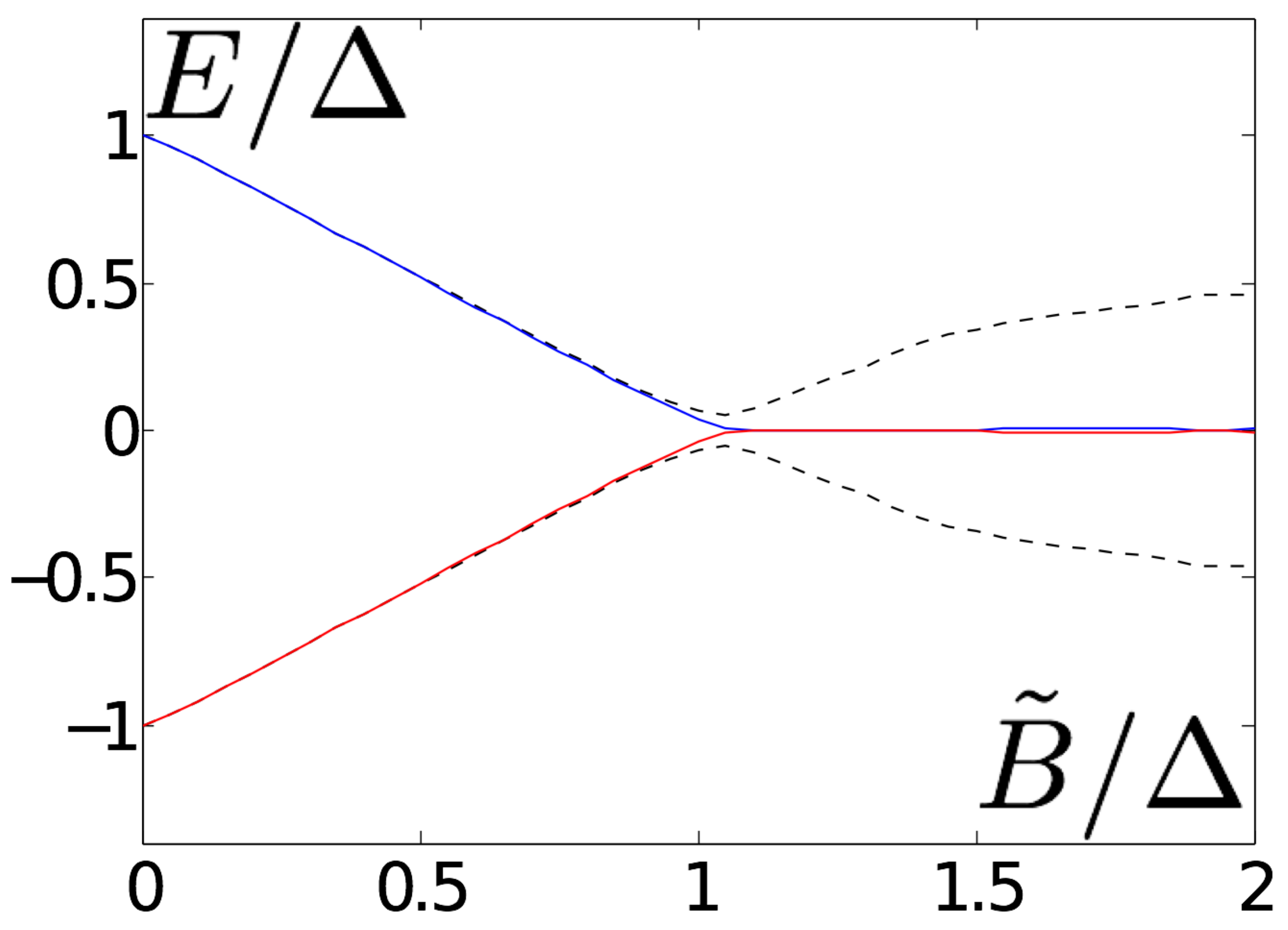}
\caption{Low-lying spectrum of the wire loop in Fig.~\ref{sche} with $r_x:r_y:r_z=6:3:1.5$, the separation of the magnets is $2R$. Left: Two positive and two negative energy states meet at $E=0$ at the critical Zeeman field, marking a topological phase transition. For a finite field interval there exists two (near) zero-energy states, corresponding to the Majorana end states, separated by a gap from the next positive and negative energy states. The parameters of the system are $N=200$, $L=20\,l_0$, $\tilde{\mu}=0$ and $R:r_x=7.5:6$. Right: same as left but for $R:r_x=8.5:6$ and $\tilde{\mu}=-0.3\Delta$. }
\label{spec1}
\end{figure}    

To study topological properties of general wire geometries we resort to numerical calculation of the wire spectrum. Our starting point is a tight-binding representation of Eq.~(\ref{h3}) with $N$ lattice sites,  
\begin{align} \label{tb}
&H=-\sum_{\langle i, j \rangle} ( t_{ij}\,\Psi_i^{\dagger}\tau_z\Psi_j+\mathrm{h.c})\nonumber\\
&+\sum_{i }(-\tilde{\mu}  \Psi_i^{\dagger}\tau_z\Psi_i+ \Psi_i^{\dagger}\vec{B}_i\cdot\vec{\sigma}\Psi_i )+\sum_{i } \Psi_i^{\dagger}\vec{\Delta}_i\cdot\vec{\tau}\Psi_i,
\end{align}
where Û\Psi_iÛ is a four-component spinor at lattice site ÛiÛ and ÛaÛ is the lattice constant. The nearest-neigbour hopping proportional to $t_{ij}$,  chemical potential $\tilde{\mu}$ and the Zeeman splitting $\vec{B}_i$ implement the first three terms on the right-hand side of Eq.~(\ref{h3}) and the last term corresponds to the superconducting order parameter $\vec{\Delta}_i$. The mapping of the continuous model (\ref{h3}) to the lattice model (\ref{tb}) implies identification $\frac{\hbar^2}{2m}=|t_{ij}|^2a$ and $\mu=\tilde{\mu}-2|t_{ij}|$. The hopping amplitude $t_{ij}$ becomes complex valued in loops enclosing a finite magnetic flux. In the following we adopt units where all lengths are given in the units of $l_0=\frac{\hbar}{(2m\Delta)^{\frac{1}{2}}}$ which yields $l_0=130$ nm for parameters $m=0.05\,m_e$ and $\Delta/k_{\mathrm{B}}=1$ K. In the following we assume that the magnets are uniformly magnetized and evaluate the stray field from the exact expression found in Ref.~\cite{engel}. The shape and strength of the field depends on the aspect ratios and remanent magnetization of the magnets and the measures of the wire. Stray fields from Fe and Co based magnets may take values of the order of 1T which determines the Zeeman energy and sets the scale for the operation temperature as discussed below.

The spectrum of the loop depicted in Fig.~\ref{sche} is illustrated in Fig.~\ref{spec1} as a function of the Zeeman splitting. The Zeeman energy is characterized by the energy scale $\tilde{B}=\frac{1}{2}g\mu_B\frac{\mu_0 M_r}{4\pi}$, where $g$ is the Lande $g$-factor, $\mu_B$ is the Bohr magneton, $\mu_0$ is the vacuum permeability and $M_r$ is the remanent magnetization of the magnets.  We have assumed that the nanowire is completely cut off at the weak link. In the topological phase we expect to find two (nearly) zero-energy MBS located at the ends determined by the weak link. As illustrated in  Fig.~\ref{spec1}, the bulk gap closes at a critical field above which there exists a pair of mid-gap states pinned to the zero energy. The wavefunctions of the mid-gap states are given by $\gamma_{1/2}=\frac{1}{\sqrt{2}}(\gamma_L\pm i\gamma_R)$  where  functions $\gamma_{L/R}=(u_{\uparrow}^{L/R}, u_{\downarrow}^{L/R},v^{ L/R}_{\downarrow},-v^{ L/R}_{\uparrow} )^T$ are localized to the left and right end of the wire and satisfy the Majorana condition $u_{\uparrow}^{L/R}=(v^{ L/R}_{\uparrow})^*$, $u_{\downarrow}^{L/R}=(v^{ L/R}_{\downarrow})^*$.  However, there is some notable differences compared to the Rashba wire model. Due to the strong spatial dependence of the magnetic field, the critical value marking the onset of the topological phase transition depends on the geometry of the system.  Also, as indicated in Fig.~\ref{spec1} (left), at sufficiently strong Zeeman fields the mid-gap states acquire finite splitting. This splitting is oscillatory in nature with increasing amplitude as the field. Topological gaps separating the MBS from the rest of the states may be a significant fraction of the induced gap $\Delta$ at vanishing magnetic field. The width of the topological phase as a function of chemical potential is comparable to $\Delta$. 

\begin{figure}
%\centering
\includegraphics[width=0.45\columnwidth, clip=true]{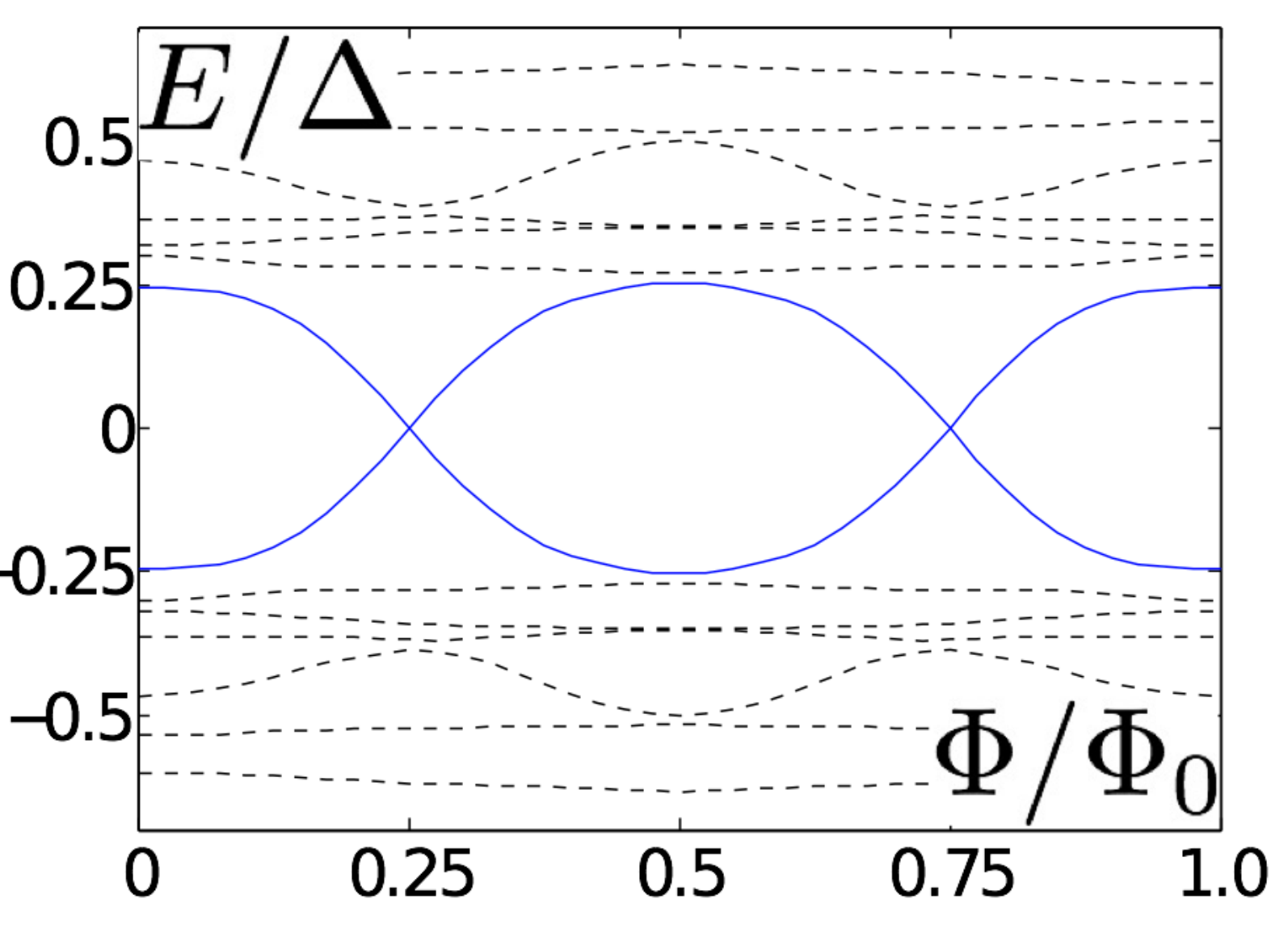}
\includegraphics[width=0.45\columnwidth ]{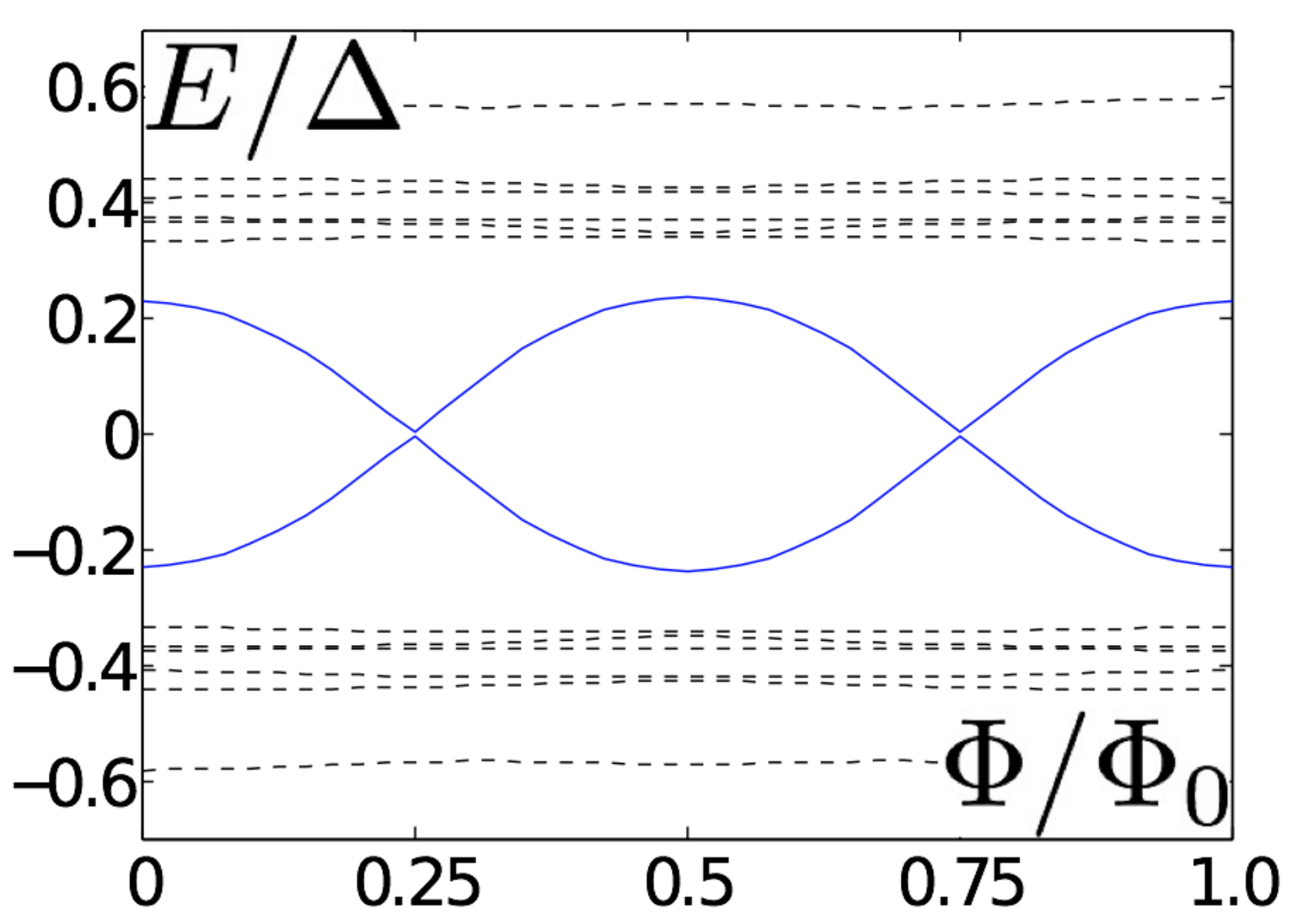}
\caption{Left: Spectrum of a SQUID loop. Solid lines represent the two MBS at the junction. The arrangement of magnets is identical to Fig.~\ref{spec1}, the other parameters are  $N=200$, $L=20l_0$ and  $\tilde{B}=\Delta$,  $\tilde{\mu}=0$ and $R:r_x=7.5:6$.  Right: Same as left but for $\tilde{B}=1.5~\Delta$, $\tilde{\mu}=-0.4\,\Delta$,  $R:r_x=8.5:6$ and lower transparency $|t_0/t_{ij}|=0.9$, where $t_0$ is the hopping element at the junction.}
\label{squid}
\end{figure}

\emph{Majorana devices}-- Now we study the properties of Josephson devices depicted in Fig.~\ref{dev}. Here we concentrate on the Andeev spectrum and neglect charging effects that could become important in mesoscopic devices \cite{vanheck}. The setup in Fig.~\ref{dev} a) has a similar structure with the system that was studied above, with the exception that the loop is enclosing a magnetic flux $\Phi$ and the weak link interrupts only the superconductor so the electrons in the nanowire may circle around. This device then can be operated as an RF SQUID. Mathematically the situation can be modelled by allowing a complex hopping amplitude $t_{ij}e^{i\Phi/\Phi_0}$ over the weak link, where $\Phi_0=\frac{h}{e}$. When the enclosed flux $\Phi$ is a multiple of $\Phi_0/4$ the junction hosts a pair of degenerate Majorana states as illustrated in Fig.~\ref{squid}. In this geometry the degeneracy is protected by the fermion parity, a fact that is not even affected by an appreciable overlap of Majorana wavefunctions through the interior of the loop \cite{pientka}. When the transparency of the junction is reduced, the energies of the MBS do not cover the whole gap but still exhibit the crossing characteristic for a topological junction.  The studied device is notably different from Josephson junctions and SQUIDs constructed from straight Rashba wire segments that generically exhibit an interplay of at least four MBS \cite{pikulin}. As a consequence, the RF SQUID geometry should display more robust $\Phi_0$-periodic features.   

\begin{figure}
%\centering
\includegraphics[width=0.45\columnwidth, clip=true]{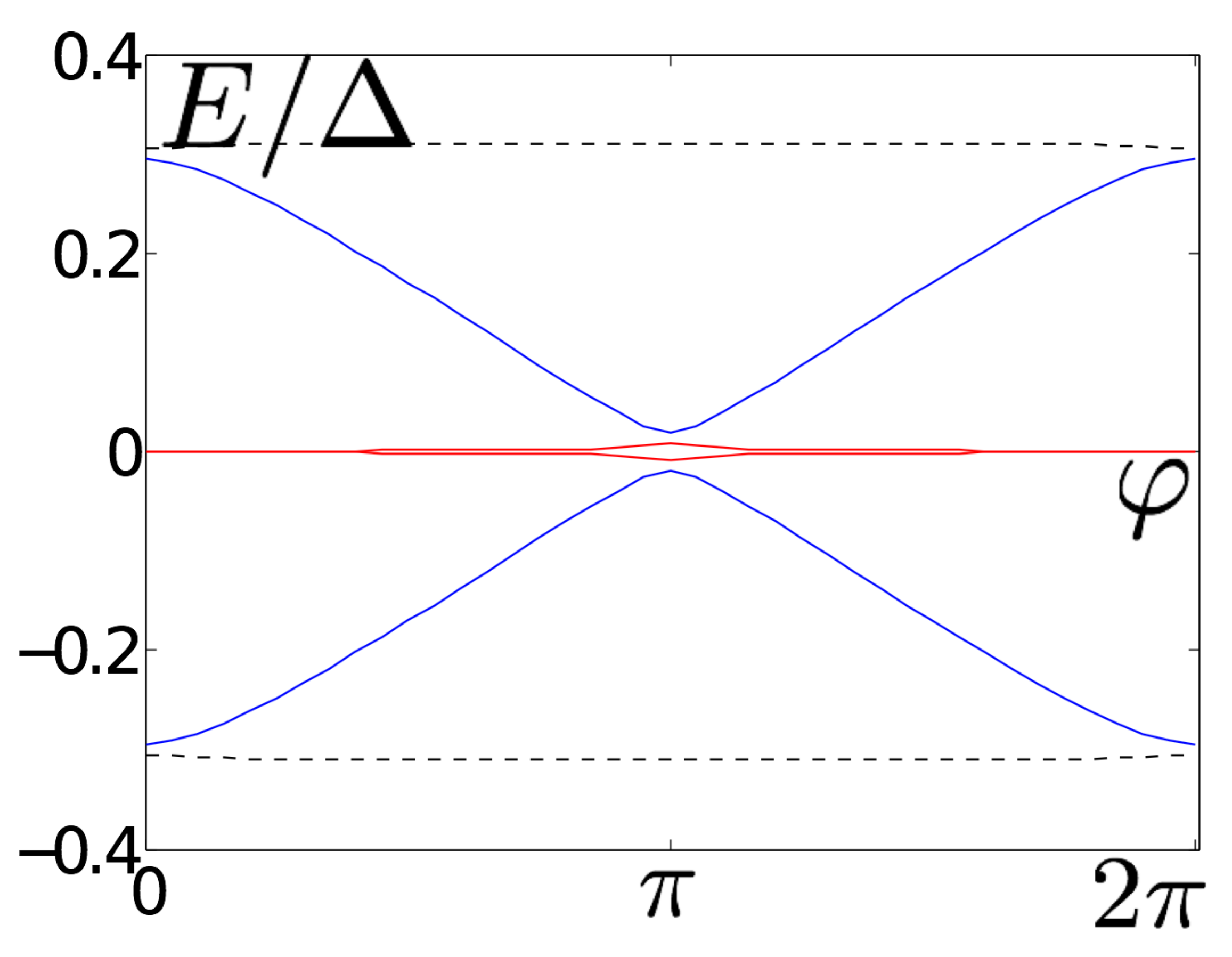}
\includegraphics[width=0.45\columnwidth, clip=true]{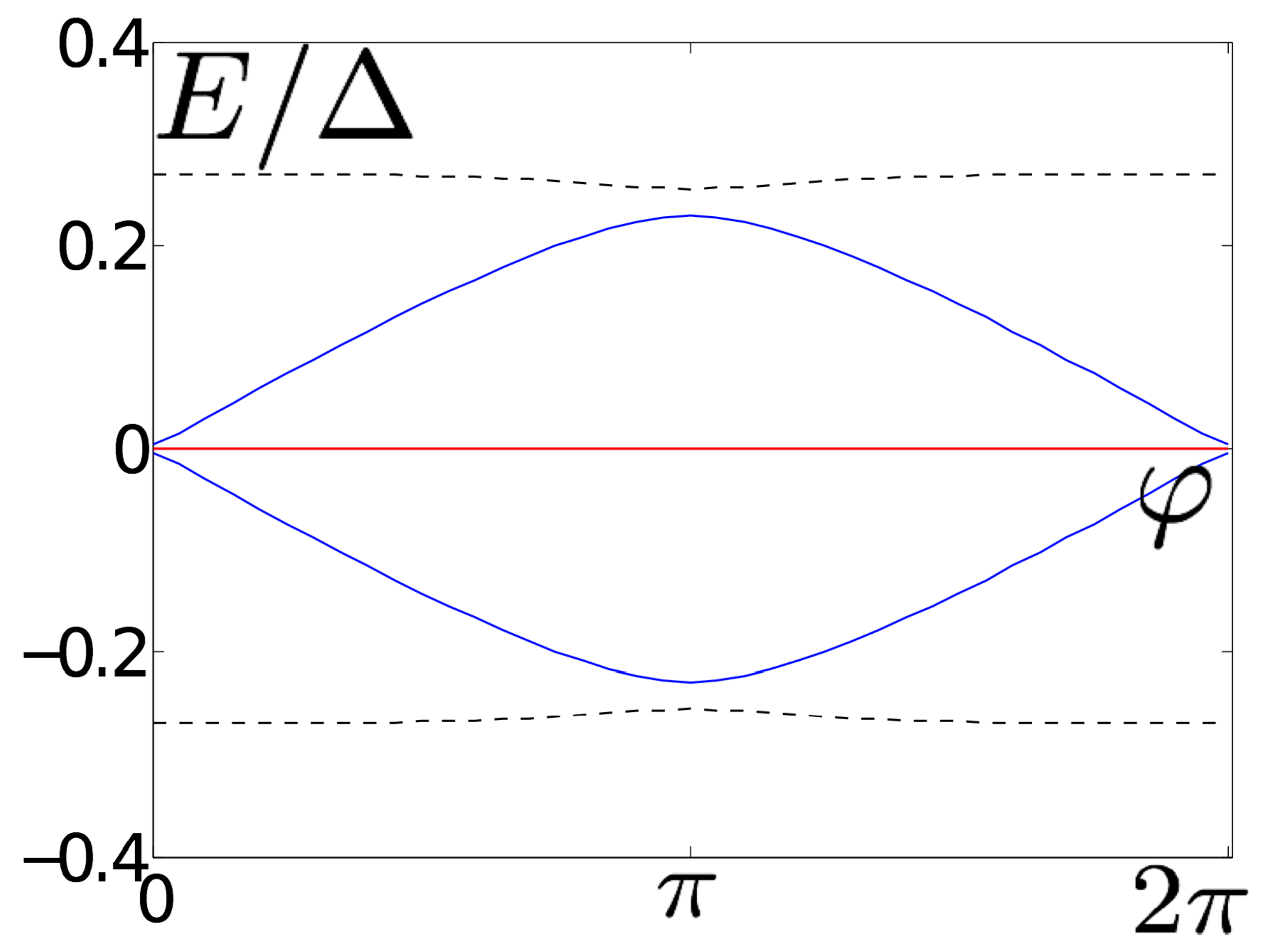}
\caption{Left: Spectrum of  the wire in Fig.~\ref{dev} b). The magnets have aspect ratio $r_x:r_y:r_z=6:3:1.5$ and are displaced by $2R$ where R is the radius of one circular arc, the other parameters are $N=200$, $L=45l_0$ and  $\tilde{B}=1.5\,\Delta$,  $\tilde{\mu}=-0.8\,\Delta$ and $R:r_x=4:3$. The solid lines represent the Majorana states at the ends (flat lines) and on the junction, the dashed lines represent the closest excited state. Right: Spectrum of the topological $\pi$-junction in Fig.~\ref{dev} c). The arrangement of magnets is as in the left and the other parameters are  $N=200$, $L=45l_0$ and  $\tilde{B}=1.5\,\Delta$,  $\tilde{\mu}=0$ and $R:r_x=4:3$.}
\label{junc}
\end{figure}

The device in Fig.~\ref{dev} b) consists of two circular arcs encircling magnets. The  wire is interrupted by a weak link in the middle and connected to separate superconducting reservoirs. The spectrum of  the setup is illustrated in Fig.~\ref{junc} (left) and exhibits a typical Majorana Josephson junction behaviour \cite{kitaev1, oreg}. At phase difference $\phi=\pi$ the junction hosts two nearly degenerate Majorana states which give rise to a weakly avoided crossing of two Andreev levels \cite{kitaev1, pikulin, sanjose} . The levels become degenerate when the overlap between the Majorana states at the junction and at the ends of the wire vanishes. The weak avoided crossing of two Andreev states is a characteristic property of topological Josephson junctions and gives rise to an effective $4\pi$ periodic Josephson effect in suitable nonequilibrium circumstances.        
An interesting situation arises in the geometry of Fig.~\ref{dev} c), where a weak link is formed between circular wire segments forming an s-shaped junction. At the centre of the wire the rotation direction of the Zeeman field is inverted, corresponding to a situation where the effective Rashba constant would change its sign. This type of junction was recently introduced in Ref.~\cite{ojanen1} and identified as a topological $\pi$ Josephson junction. As shown in Fig.~\ref{junc} (right), the spectrum of the MBS at the junction is qualitatively shifted by $\pi$ compared to Fig.~\ref{dev} b) and thus exhibits a weak avoided crossing at phase difference $\phi=0$. The possibility of two degenerate MBS at zero phase difference implies existence of more than two distinct topological phases enabled by chiral symmetry in one-dimensional systems \cite{schnyder, ojanen1, diez,  vayrynen}.   Since supercurrent through the junction is given by the phase gradient of the populated levels, the maximum supercurrent through a topological $\pi$ junction achieved, remarkably, near the vanishing phase difference $\phi=0$ (mod $2\pi$).

Effects of disorder and various unideal features on MBS present in real systems have been studied widely in the Rashba wire model and qualitatively similar behaviour is expected for the studied systems. Above we have discussed physical properties when the intrinsic spin-orbit coupling, depending on the details of the setup, can be neglected. However, most of the presented conclusions remain qualitatively unaltered in the presence of weak Rashba coupling $l_{\mathrm{so}}\gtrsim5l_{0}$ where $l_{\mathrm{so}}=\frac{\hbar}{m\alpha}$.  In this case the spin-orbit term introduces a slight shift of topological degeneracy points. The notable exception is the topological $\pi$-effect which is strongly affected if $l_{\mathrm{so}}\gtrsim 10 l_{0}$ is not satisfied.   

Nanowires made out of InSb and InAs  are currently leading candidates to exhibit MBS. Producing Majorana networks by growing and placing nanowires individually is not efficient, thus motivating alternative approaches. In Fig.~\ref{dev} d) we have shown one possible structure capable of supporting a large number of pairs of MBS. Topological properties of each loop is similar to the studied structure in Fig.~\ref{sche}. This type of array is an example of a potentially useful building block in Majorana devices and does not require application of external magnetic field. Two loops could be coupled, for example, through their magnetic fluxes. It is also straightforward to imagine more complex unit blocks with more MBS. Since magnetization of all the magnets is parallel, it is simple to tune or reinforce it by external magnetic field parallel to the array.

\emph{Discussion}--The proximity superconductivity in semiconducting wire loops could be arranged along the lines of Ref.~\cite{fornieri} which demonstrated a fabrication required structure without magnets. One possibility to realize Josephson devices is to place the superconducting metal on top of the wire like in Ref.~\cite{rokhinson}. It has also been demonstrated that magnets of required dimensions can be fabricated in the proximity of complicated structures by presently existing methods \cite{deon}.  Wires could be made of semiconductor materials weak spin-orbit coupling, though most promising candidates of high $g$-factor and low effective mass also commonly have interesting spin-orbit properties.  The candidate materials include, for example, Ga$_x$In$_{1-x}$As, GaSb, GaInSb, InSb and InAs all of which may have large $g$-factors.  Fabrication of InAs 2DEG proximity coupled to superconductors is well-known technology making it one of the most promising candidate material. It is not necessary to work with single-channel wires but the ability to tune the electron density by a gate voltage is needed to ensure that the nontrivial phase achieved. 

If the magnets are made of Fe or Co- based materials, remanent magnetization could be $\mu_0M_r\approx 1.8\,$T or higher, translating to the characteristic Zeeman energy 
$\tilde{B}=\frac{1}{2}g\mu_B\frac{\mu_0 M_r}{4\pi}\approx k_{\mathrm{B}}0.7\,$K for InAs parameters $g=15$, $m=0.023\,m_0$. As Figs.~\ref{spec1}, \ref{squid} show, the induced gap $\Delta$ in the topological regime should comparable to $\tilde{B}$. In this regime the topological gap separating MBS from other states is $\sim 0.3\Delta/k_\mathrm{B}\approx 200\,$mK so the operating temperature should be significantly smaller than this. For InSb the effective mass is $m=0.016\,m_0$ and the $g$-factor could reach 50, enabling topological gaps of the order of 1K. Employing InAs parameters, the geometrical measures of the setup in Fig.~\ref{sche} corresponding to parameters in Figs.~\ref{spec1} (left) and \ref{squid} (left) are $L=2\pi R=3.3\,\mu$m, so the loop diameter is $2R=1.1\,\mu$m and the magnets have measures $r_x=850$ nm $r_y=420$ nm, $r_z=210$ nm.  

The observable signatures of topological junctions arise from properties of mid-gap MBS. The fractional Josephson effect, manifesting as an anomalous Shapiro step, has been seen in experiments \cite{rokhinson}. The junction in Fig.~\ref{dev} b) should exhibit this type of behaviour. In the RF SQUID geometry there is only two MBS in the system that could be observed through a telegraph noise of current in the loop associated with quasiparticle number fluctuations in the junction \cite{fu2} and through a $\Phi_0$-periodic component of equilibrium supercurrent \cite{pientka}.    

\emph{Summary}--In this work we proposed a method to realize Majorana devices by fabricating superconducting loops in the vicinity of permanent magnets. In these systems it is possible achieve Majorana devices with weak or no spin-orbit coupling and without external magnetic field. In addition, these systems enable  fabrication of devices that are unfeasible by employing straight wire segments. The topological RF SQUID has the simplest device geometry and can be realized by currently existing technology. Magnetic arrays could also enable controlled fabrication of Majorana arrays potentially useful in quantum information applications. 

The author would like to thank Pauli Virtanen and Karsten Flensberg for discussions and acknowledge the Academy of Finland for support.

\end{document}